\documentstyle[twoside,fleqn,espcrc2]{article}

\newcommand{\eea}{\end{eqnarray}}
\def\beq{\begin{equation}}
\def\eeq{\end{equation}}
\def\beqa{\begin{eqnarray}}
\def\eeqa{\end{eqnarray}}
\newcommand{\be}{\begin{equation}}
\newcommand{\eq}{\end{equation}}

\newcommand{\AmS}{{\protect\the\textfont2
  A\kern-.1667em\lower.5ex\hbox{M}\kern-.125emS}}

\title{$N=1$ Dual String Pairs and their Modular Superpotentials}

\author{Dieter L\"ust\address{Humboldt-Universit\"at, Institut f\"ur Physik, \\ 
        Invalidenstra{\ss}e 110, 10115 Berlin, Germany}%
        \thanks{The work is supported by the Deutsche 
        Forschungsgemeinschaft (DFG)
and by the European Commission TMR programme ERBFMRX-CT96-0045.}}
        
\begin{document}

\begin{abstract}
We review the duality between
heterotic and F--theory string vacua with $N=1$
space-time supersymmetry in eight, six and four dimensions.
In particular, we discuss
two chains of four-dimensional
F--theory/heterotic dual string pairs, where F--theory is compactified
on certain elliptic Calabi-Yau fourfolds, and the dual heterotic vacua are 
given
by compactifications on  elliptic
Calabi-Yau threefolds plus the specification
of the $E_8\times E_8$ gauge bundles.  
We show that   the massless spectra of the dual pairs agree
by using, for one chain of models, an index formula
to count the heterotic bundle moduli and 
determine the dual F--theory spectra from the Hodge numbers
of the fourfolds  and of the type IIB base spaces.
Moreover as a further
check, we demonstrate that for one particular heterotic/F--theory dual
pair the $N=1$ superpotentials are the same.

\end{abstract}

\maketitle

\section{Introduction}

Several types of strong-weak coupling duality symmetries in string theories
were explored and established during recent years 
\cite{sdual,witt,KV,dualities}. 
As a result of these non-perturbative investigations
it seems nowadays clear that all five known
consistent string constructions, the two heterotic strings, the type IIA,B
superstrings and the type I superstring,
are related by some kind of duality transformation such that they are all
equivalent and on equal footing. 
Moreover it was a very important and fascinating observation \cite{witt}
that a strongly coupled string theory, after including non-perturbative
states, may look like a theory which lives in higher dimensions. Specifically,
the Dirichlet (D) 0-branes of the IIA superstring behave in the same way as
the Kaluza-Klein states of 11-dimenensional supergravity, and in fact 
there is very strong evidence
that    the
type IIA string in ten dimensions is dual to 11-dimensional supergravity
compactified on a circle of radius $R$.
The full quantum version of 11-dimensional
supergravity is now called $M$-theory; $M$-theory on circle of radius $R$
is supposed to 
describe the full strongly coupled type IIA superstring.
However a fundamental definition of $M$-theory
is up to now  not clear. One promising attempt to define $M$-theory
is given by a Hamiltonian theory of non-commuting matrices, known as
M(atrix)-theory \cite{matrix}.

In \cite{V} Vafa made the proposal that non-perturbative string physics can lead
to dimensions even beyond 11, namely he
conjectured
 that the IIB superstring theory should
be regarded as the toroidal compactification of twelve--dimensional
F--theory. This would lead to supergravity theories with two time and ten
spatial directions which are hardly understood (for progress along these
lines see \cite{barskounn}).
However,
if one is slightly less ambitious and does not want to enter
the enormeous difficulties in formulating a consistent 12-dimensional
theory, $F$-theory can be regarded as to provide
new non-perturabtive type IIB vacua
on D--manifolds  
in which the complexified coupling varies over the
internal space. These compactifications then have a beautiful
geometric interpretation as compactifications of $F$-theory on
elliptically fibred manifolds, where the fibre encodes the behaviour of
the coupling, the base is the D--manifold, and the points where
the fibre degenerates specifies the positions of the D 7-branes 
in it. Compactifications of F--theory on elliptic Calabi--Yau
twofolds (the K3), threefolds and fourfolds can be argued \cite{V} to be
dual to certain heterotic string theories in 8,6 and 4 dimensions and
have provided new insights into the relation between geometric 
singularities and perturbative as well as non--perturbative 
gauge symmetry enhancement and into the structure of moduli spaces.

Let us consider the duality between the heterotic string 
and F--theory 
in slightly more detail. To compactify F--theory
down to $D$ dimensions we are dealing with a Calabi-Yau space of complex
dimension $6-{D\over 2}$, called $X^{6-{D\over 2}}$,
which is elliptically fibred over a complex base $B^{5-{D\over 2}}$
of complex dimension $5-{D\over 2}$.
This non-Calabi-Yau space $B^{5-{D\over 2}}$
is the type IIB compactification space from ten to $D$
dimensions. Moreover one supposes that 
$B^{5-{D\over 2}}$ is rationally ruled, 
i.e. there exists a fibration
$B^{5-{D\over 2}}\rightarrow \tilde B^{4-{D\over 2}}$ 
with $P^1$ fibers, because one has to assume 
the  $X^{4-{D\over 2}}$ to be a $K3$ fibration
over the two dimensional surface $\tilde B^{4-{D\over 2}}$. So,
we have the following fibration structure for $X^{6-{D\over 2}}$:
\begin{equation}
X^{6-{D\over 2}}\rightarrow_{T^2}B^{5-{D\over 2}}\rightarrow_{P^1}\tilde B^{4-{D
\over 2}}.\label{fourfibr}
\end{equation}
Then, using the eight-dimensional
duality among F--theory on an elliptic $K3$ and the heterotic string
on $T^2$, one derives by an adiabatic argument that the 
dual heterotic vacua are given by
Calabi-Yau compactifications on a threefold 
$Z^{5-{D\over 2}}$, which is an elliptic fibration
over the same surface $\tilde B^{4-{D\over 2}}$, 
\begin{equation}
Z^{5-{D\over 2}}\rightarrow_{T^2}\tilde B^{5-{D\over 2}},\label{hetfibr}
\end{equation}
and where the choice of an $E_8\times E_8$ gauge vector
bundle $V_1\times V_2$ has
to be specified in order to match all F--theory fibration data.

In this paper we will consider $D$-dimensional dual pairs with $N=1$
space-time supersymmetry. Specifically, in the next section we briefly discuss
$D=8$, i.e.   F--theory on $K3$
being dual to the heterotic string on $T^2$, then in section three we 
review
 $D=6$ F--theory on  Calabi-Yau
threefolds $X^3$ \cite{MV} which is dual to the heterotic string 
on $K3$ and finally,
the main part of our paper, we discuss $D=4$
F--theory on fourfolds $X^4$ \cite{fourfolds,sethi,fmw,cl,acl}
and
their heterotic duals on Calabi-Yau threefolds $Z^3$.
On the F--theory side
as well as on the heterotic side the geometry of the 
considered manifolds relies on del Pezzo surfaces. 
To establish the four-dimensional
F--theory/heterotic duality 
we match the massless spectra \cite{cl,acl}
by using, for one chain of models, an index formula \cite{fmw}
to count the heterotic bundle moduli and 
determine the dual $F$-theory spectra from the Hodge numbers
of the fourfolds $X^4$ and of the type IIB base spaces.
 Moreover, for models
which we construct by a ${\bf Z}_2$ modding of dual 4D pairs with $N=2$
space-time supersymmetry we can show that the F--theory/heterotic
$N=1$ superpotentials agree \cite{cl}.
Here the superpotentials, which were computed on
the  F--theory side in \cite{wit1,dgw}, are given by certain modular
functions, similar to the prepotentials which appear in $N=2$ dual
typeII/heterotic string pairs \cite{KV,hetprep}.

\section{Eight Dimensions}

The heterotic string compactified on $T^2$ to eight dimensions is well
known.
The heterotic moduli are given by 18 complex parameters which divide
themselves into the K\"ahler class plus complex structure of $T^2$,
called $T$ and $U$, and 16 Wilson line moduli, which determine
the $E_8\times E_8$ gauge bundle, i.e. the unbroken gauge group in eight
dimensions. The moduli space of the 18 complex moduli is given by the
well known Narain coset \cite{narain}
\begin{equation}
{\cal M}={SO(2,18)\over SO(2)\times SO(18)},\label{classmod}
\end{equation}
where the discrete $T$-duality group $SO(2,18,{\bf Z})$ has still to be
modded out. At special points/loci the generic Abelian gauge group
$U(1)^{18}(\times U(1)^2)$ can be enhanced; e.g. at the line $T=U$ there
is an $SU(2)$ gauge symmetry enhancement and for tuned Wilson lines one
recovers the full $E_8\times E_8$ gauge symmetry.

The dual type IIB compactification to eight dimensions on the space $B^1=P^1$
with complex coordinate $z$ is characterized by a complex dilaton field
$\tau$, which varies holomorphically over $z$, i.e. $\tau=\tau(z)$.
To fullfil the corresponding equations of motion a non-perturbative
background of 24 7-branes has to be turned on. Then
Vafa has argued \cite{V}
that the heterotic string compactified
on a two--torus in the presence of Wilson lines 
is dual to F--theory compactified on the family 
\be
y^2 = x^3 + f_8(z) x + g_{12}(z)
\label{K3withWL}
\eq
of elliptic K3 surfaces, where $f_{8}(z)$, $g_{12}(z)$ are polynomials
of order $8,12$ respectively.
In particular F--theory on the two parameter subfamily
\be
y^2 = x^3 + \alpha z^4 x + (z^5 + \beta z^6 + z^7)
\label{K3withoutWL}
\eq
of K3's with $E_8$ singularities at $z=0,\infty$
is dual to the heterotic theory with Wilson lines switched off \cite{MV}.
Therefore there must exist a map which relates the complex structure and
K\"ahler moduli $U$ and $T$ of the torus on which the heterotic theory
is compactified to the two
complex structure moduli $\alpha$ and $\beta$ in
(\ref{K3withoutWL}). This map can be worked out explicitly, and one gets
\cite{cclm}
\beqa
j(iT) j(iU) &=&-1728^2\frac{\alpha^3}{27},\nonumber\\
(j(iT) - 1728)(j(iU) - 1728)&=&\;\; 1728^2\frac{\beta^2}{4}.
\label{Resulta}
\eeqa
Let us  remark that our result can be regarded as the two-parameter
generalization of the one-parameter torus in the
well-known
Weierstrass form ($g_2=\frac{4}{3}\pi^4 E_4$, $g_3=
\frac{8}{27}\pi^6 E_6$)
\be
y^2 = 4 x^3 - g_2(\tau) x - g_3(\tau).
\eq
As a check of the result eq.(\ref{Resulta})
we compute the discriminant of the $K3$ eq.(\ref{K3withoutWL}):
\be
\Delta^{(K3)} = \left( \alpha^3 + \frac{27}{4} \beta^2 + 27 \right)^2 
- 27^2 \beta^2,
\label{K3discriminant}
\eq
It is easy to see that $\Delta^{(K3)}=0$ at the line $T=U$ of $SU(2)$
gauge symmetry enhancement. So the $K3$ becomes singular at this locus, and
as long as both $\alpha$ and $\beta$ are not zero,
the singularity is of type $A_1$.

\section{Six Dimensions}

Here we consider first the heterotic string on $K3$. The heterotic gauge
bundle is specified by the numbers $(n_1,n_2)$ of $E_8\times E_8$ instantons
turned on. 
In addition one can consider also the 
non-perturbative background of $n_5$ heterotic 5-branes \cite{seiwi}.
The choice of the three integers $n_1$, $n_2$ and $n_5$ is restricted
by the following Green-Schwarz anomaly matching condition
\begin{equation}
n_1+n_2+n_5=24.\label{gsmatch5}
\end{equation}
The massless
spectrum of the resulting six-dimensional $N=1$ supergravity is then
determined as follows. Besides the supergravity multiplet, we have first
$N_H$,
\be
N_H=20+\dim_Q{\cal M}_{\rm inst}+n_5,\label{nh}
\eq
hyper multiplet moduli fields which parametrize the Higgs branch of the
theory. Here the 20 counts the number of $K3$ moduli, the second term
in eq.(\ref{nh}) denotes the dimension 
 of the quaternionic instanton moduli space of the embedded
$E_8$ instantons, and the last
term arises, since the position of each five brane on $K3$
is parametrized by a hypermultiplet.
The dilaton $\phi$ plus the self-dual antisymmetric tensor field
$B_{\mu\nu}^+$ are members of a tensor multiplet. In addition,
considering the non-perturbative contribution
of heterotic 5-branes, there are
  additional tensor multiplets in
the massless spectrum \cite{seiwi}, 
since on the world sheet theory of the 5-brane
lives a massless tensor field. Hence the total number $N_T$ of six-dimensional
tensor fields is
\begin{equation}
N_T=1+n_5.\label{nt}
\end{equation}
Thus the Coulomb branch of the
six-dimensional theory is characterized by 
a real
$(1+n_5)$-dimensional moduli space, parametrized by the scalar field vev's
of the tensor multiplets.
Finally, one gets massless vector multiplets associated to the part
of the gauge group $G_1\times G_2$ of rank $r(V)$
which is left unbroken by the instantons.
So $N_V=\dim(G_1\times G_2)$.
If $n_a\geq 10$ ($a=1,2$) the gauge group $E_8$ is completely broken at a
generic point in the hypermultiplet moduli space, whereas for smaller
values of $n_a$, there will be always an unbroken non-Abelian gauge symmetry.
Note that at special loci in the hypermultiplet moduli space 
the instantons may fit into a smaller subgroup of
$E_8$ and the gauge group $G_a$
can be enhanced, which is just the (reverse) Higgs effect in field theory.

Let us now discuss F--theory compactifications on  
Calabi-Yau threefolds $X^3$ which are dual to the heterotic $K3$
compactifications just discussed before.
$X^3$ has to be an elliptic fibration over a base $B^2$, i.e. $X^3\rightarrow
B^2$. As explained in \cite{MV} the spectrum of six-dimensional tensor,
(Abelian) vector and hypermultiplets can be computed from the Hodge numbers of
$X^3$ together with the Hodge numbers of $B^2$ in the following way:
\begin{equation}
N_H=h^{(2,1)}(X^3)+1,\label{nohyper}
\end{equation}
\begin{equation}
N_T=h^{(1,1)}(B^2)-1,\label{notensor}
\end{equation}
\begin{equation}
r(V)=h^{(1,1)}(X^3)-h^{(1,1)}(B^2)-1.\label{novector}
\end{equation}
These numbers have to match up the corresponding numbers on the
heterotic side for a given dual string pair and imply in particular a
mapping of the heterotic data, $(n_1,n_2,n_5)$ on the F--theory Hodge numbers.

Let us discuss in more detail a particular class of dual models.
 On the heterotic
side, they are characterized by the absence of five-branes $n_5=0$; so there
are no non-perturbative tensor fields, i.e. $N_T=1$.
The anomaly equation (\ref{gsmatch5}) then implies that $n_1=12+k$ and 
$n_2=12-k$ ($k\geq0$). Assuming maximal Higgsing the spectrum of vector
and hypermultiplets is then given for the following choices of $k$ as:
\beqa
\begin{tabular}{|c||c|c|} \hline
  $k$     & $G_2$  & $\dim{\cal M}_{\rm inst}$ \\ \hline\hline
0  &   1 & 224   \\ \hline
2 &   1 & 224   \\ \hline
4 &    $SO(8)$& 252 \\ \hline
6 & $E_6$ & 302 \\ \hline
12 & $E_8$ & 472 \\ \hline
\end{tabular}
\label{tab5}
\eeqa
($G_1$ is always completely Higgsed.)

On the F--theory side, the compactifications which are dual to these 
perturbative heterotic string vacua are given by a class of elliptic threefolds
$X^3_k$, where the bases $B^2_k$ are given by the k-th Hirzebruch
surface $F_k$. 
These surfaces are all $P^1$ fibrations
over $P^1$, and they are distinguished by how the 
$P^1$'s are
twisted. For example, $F_0$ is just the direct product $P^1\times P^1$.
For all $F_k$,
 $h^{(1,1)}(F_k)=2$. Therefore one immediately gets
that $N_T=1$, which corresponds to the universal 
heterotic dilaton tensor multiplet
in six dimensions. 
In the following it will become very useful to describe the elliptically
fibred Calabi-Yau spaces $X^3_k$ in the Weierstrass form \cite{MV}:
\begin{eqnarray}
X^3_k:\quad 
y^2&=&x^3+\sum_{n=-4}^4f_{8-nk}(z_1)z_2^{4-n}x
\nonumber \\ &+&\sum_{n=-6}^6g_{12-nk}(z_1)
z_2^{6-n}.\label{weier}
\end{eqnarray}
Here $f_{8-nk}(z_1)$, $g_{12-nk}(z_1)$ are polynomials of degree
$8-nk$, $12-nk$ respectively, where the polynomials with negative degrees
are identically set to zero.
From this equation we see that the Calabi-Yau
threefolds
$X^3_k$ are $K3$ fibrations over $P^1_{z_1}$ with coordinate $z_1$; the 
$K3$ fibres themselves are elliptic fibrations over the $P^1_{z_2}$ with
coordinate $z_2$.
The Hodge numbers $h^{(2,1)}(X^3_k)$, 
which count the number of complex structure
deformations of $X^3_k$, are  given by the the number of parameters
of the curve (\ref{weier}) minus the number of possible reparametrizations,
which are given by 7 for $k=0,2$ and by $k+6$ for $k>2$.
One can explicitly check the Hodge numbers derived in this way 
precisely match the
heterotic spectra in table (\ref{tab5}) using eqs.
(\ref{nohyper},\ref{notensor},\ref{novector}).
The non-Abelian gauge symmetries
are determined 
by the 
singularities of the curve (\ref{weier}) and were analyzed in detail in
\cite{BIKMSV}.
For example, for $k=0,1,2$ it is easy to see that the elliptic curve
(\ref{weier}) is generically non-singular. Only tuning some parameters
of the polynomials $f$ and $g$ to special values, the curve will become
singular.
These F--theory
singularities correspond to the perturbative gauge symmetry enhancement
in the dual heterotic models.
On the other hand, for the cases $k>2$ the curve (\ref{weier}) always
contains generic singularities, since on the heterotic side the gauge
group cannot be completely Higgsed.

Upon compactification on a two-dimensional
torus $T^2$, the six-dimensional $N=1$
duality between F--theory  and the heterotic string  leads to a $N=2$ duality
between
F--theory on $X^3\times T^2$ dual to heterotic on $K3\times T^2$.
This duality is then extended \cite{V} by observing that the 
four-dimensional, $N=2$ supersymmetric heterotic string
on $K3\times T^2$ is also dual \cite{KV} to the type IIA 
superstring compactified
on the same Calabi-Yau threefold $X^3$. The number of $N=2$ 
hypermultiplets is  given by $N_H=h^{(2,1)}(X^3)+1$, and the
number of $U(1)$ $N=2$ vector multiplets
is  given by $N_V=h^{(1,1)}(X^3)$ in agreement with 
eqs.(\ref{nohyper},\ref{notensor},\ref{novector}), taking into account
that the $T^2$ compactifications leads to two addtional $U(1)$ vector fields
$T$ and $U$. 
We will later use this $N=2$ F--theory/heterotic duality to 
construct
dual pairs with $N=1$ space-time supersymmetry by a ${\bf Z}_2$ modding
procedure.

\section{Four Dimensions}

As a continuation of the six-dimensional string vacua discussed in the
previous chapter, 
$N=1$ supersymmetric in four dimensions are obtained either by compactifying
the heterotic string on a Calabi-Yau threefold $Z$ or, in the dual description,
by F--theory compactification on a fourfold $X^4$. Let us first discuss
some properties of the heterotic vacua. Additional to the choice of the
threefold $Z$ the heterotic vacuum must be further specified by a particular
choice of the gauge bundle $V_1\times V_2$, which determines the breaking
of the original gauge group $E_8\times E_8$ to some subgroup $G_1\times G_2$
and also, together with the data of $Z$, the chiral $N=1$ matter fields, which
transform non-trivially under $G_1\times G_2$. In general, the matter
field representations will be in chiral representations with respect to
the gauge group. Unlike six dimensions, where the gauge bundle can be
characterized by two integers, the instanton numbers $n_1$ and $n_2$,
the four-dimensional gauge bundle is characterized by the second Chern
classes of $V_a$, called $c_2(V_a)$.
In analogy to the Green-Schwarz condition eq.(\ref{gsmatch5}) in six
dimensions, there is a anomaly constraint on the four-dimensional gauge
bundle including non-perturbative 5-brane \cite{fmw}:
\begin{equation}
c_2(V_{1})+c_2(V_{2})+n_5=c_2(Z).\label{4danomaly}
\end{equation}
For perturbative heterotic vacua with $n_5=0$, 
this anomaly condition can be always satisfied by the socalled standard
embedding of the spin connection into the gauge fields, i.e. identifying
the $SU(3)$ holonomy group of $Z$ with a subgroup of one $E_8$. This choice
leads to the class of socalled (2,2) superconformal $N=1$ heterotic vacua with
gauge group $E_6\times E_8$ and $h^{(1,1)}(Z)$ chiral matter fields in
the ${\bf 27}$ representation plus $h^{(2,1)}(Z)$ in the ${\bf{\bar{27}}}$
representation of $E_6$. However this is by far not the only possible choice,
and a general (perturbative) heterotic vacuum is given by a (0,2) superconformal
field theory. In fact, all models with a F--theory dual we know so far, are
(0,2) compactifications for which the gauge bundle is often very hard
to be explicitly constructed.

The moduli of a general (0,2) heterotic Calabi-Yau compactification count
first the possible deformations of $Z$ and, second, also the possible 
deformations of the gauge bundle $V_1\times V_2$ (corresponding to the
instanton moduli space in six dimensions).
So the number of gauge singlet moduli, chiral plus antichiral fields is given by
\begin{equation}
N_C=h^{(1,1)}(Z)+h^{(2,1)}(Z)+n_{\rm bundle},\label{nchiral}
\end{equation}
where $n_{\rm bundle}$ counts the bundle deformations. The number of $N=1$
vector multiplets is simply given by the dimensions of the unbroken
gauge group, $N_V=\dim(G_1\times G_2)$. 
In addition there will in general matter,
charged under the gauge group. Like in six dimensions, for special
values of the chiral moduli fields, the gauge group may be extended.

As already said in the introduction, we assume that $Z$ is an elliptic
threefold, i.e. there exist the fibration $Z\rightarrow_{T^2} \tilde B^2$. 
Because of this elliptic fibration structure there exist a ${\bf Z}_2$
involution $\tau$ which acts trivially on the base $\tilde B^2$ but
acts as multiplication by -1 on the elliptic fibres \cite{fmw}.
It follows that the gauge bundle moduli divide themselves into even and odd
chiral superfields under $\tau$: $n_{\rm bundle}=n_e+n_o$.
On the other hand, the difference $n_e-n_o$ can be computed from an index
calculation \cite{fmw} as follows. Namely consider the following index
\begin{equation}
I=-\sum_{i=0}^3(-1)^{i}\dim H_e^i, \label{index}
\end{equation}
where ${\rm Tr}_{H^i(Z,{\rm ad}(V))}{1+\tau
\over 2}=\dim H_e^i$.
Here $\dim H_e^1$ is $n_e$, $\dim H_e^2$ is $n_o$. Moreover,
$\dim H_e^0$ and $\dim H_e^3$ are the numbers of unbroken gauge generators
that are even or odd under $\tau$. So in case the gauge group is completely
broken we simply get
\begin{equation}
I=n_e-n_o.\label{simpleindex}
\end{equation}

Now consider the dual compactifications of F--theory on $X^4$.
As discussed in the introduction, $X^4$ is assumed to have the following
fibration structure:
$X^4\rightarrow_{T^2}B^3\rightarrow_{P^1}\tilde B^2$. The cohomology of $X^4$
is characterized by four Hodge numbers, namely the number of K\"ahler 
deformations $h^{(1,1)}(X^4)$, the number of complex structure
deformations $h^{(3,1)}(X^4)$ and two Hodge numbers $h^{(2,1)}(X^4)$ and
$h^{(2,2)}(X^4)$, where the last Hodge number will not play any role in
the following. As discussed in \cite{sethi}, the requirement of tadpole
cancellation demands that $n_3={\chi\over 24}$ type IIB 3-branes have to
be turned on, where $\chi$ is the Euler number of $X^4$. According
to \cite{sethi} there is the following relation
between the Hodge numbers $h^{(1,1)}(X^4)$,
$h^{(2,1)}(X^4)$ and $h^{(3,1)}(X^4)$: ${\chi\over 6}=
8+h^{(1,1)}(X^4)-h^{(2,1)}(X^4)+h^{(3,1)}(X^4)$.
One can in fact show \cite{fmw,ac}
that for a heterotic/F--theory dual string pair
the numbers of 5-branes and 3-branes coincide: $n_3=n_5$. Hence,
if the Euler number of $X^4$ is non-vanishing we are dealing with a 
non-perturbative heterotic string vacuum.

Let us now discuss the map of the fourfold data to the heterotic data, provided
by the Calabi-Yau threefold $Z$ plus the gauge bundle $V_1\times V_2$.
To start with, let us indicate the general structure of the F--theory/heterotic
correspondence. First, the number of $X^4$ K\"ahler deformation
$h^{(1,1)}(X^4)$ will correspond to the number of K\"ahler deformation
$h^{(1,1)}(Z)$ of the heterotic threefold $Z$ and will also be related
to the rank of the unbroken gauge group $r(V)$:
\begin{equation}
h^{(1,1)}(X^4)\, \leftrightarrow \, h^{(1,1)}(Z),r(V)\label{h1z1}
\end{equation}
 Second, the number of  complex
structure deformations $h^{(3,1)}(X^4)$ will be in correspondence to
the number of complex structure deformations of $Z$ plus the number of even
gauge bundle deformations:
\begin{equation}
h^{(3,1)}(X^4)\, \leftrightarrow \, h^{(2,1)}(Z)+n_e.\label{h3z2}
\end{equation}
Finally, the Hodge number $h^{(2,1)}(X^4)$ is in correspondence with the odd
bundle deformations:
\begin{equation}
h^{(2,1)}(X^4)\, \leftrightarrow \, n_o.\label{h2no}
\end{equation}

To be more precise,
 a refined analysis shows that like in the six-dimensional case
also the topological data of the base $B^2$ enter the four-dimensional
spectrum. Specifically as discussed in \cite{mohri,cl,acl},
looking at the dimensional reduction of the ten-dimensional type IIB
spectrum, the following formulas are obtained
\beqa
r(V)&=&h^{(1,1)}(X^4)-h^{(1,1)}(B^3)-1 \nonumber\\
&+&h^{(2,1)}(B^3),\label{spectrumv} 
\eeqa
\beqa
N_C&=&h^{(1,1)}(B^3)-1+h^{(2,1)}(X^4)\nonumber \\ 
&-&h^{(2,1)}(B^3)+h^{(3,1)}(X^4)
\nonumber\\
 &=&
 \frac{\chi}{6}-10+2h^{(2,1)}(X^4)-r(V).
 \label{spectrumc}
\eeqa
Note that in this formula we did not count the chiral field which 
corresponds to the dual heterotic dilaton.
Compactifying F--theory further to three dimensions on $X^4\times S^1$
(or equivalently to two dimensions on $X^4\times T^2$)
these equations are consistent with M--theory compactification on $X^4$
to three dimensions, since there the sum $r(V)+N_C$ must be independent from
the data of the base $B^2$ and is given by $r(V)+N_C=
h^{(1,1)}(X^4)+h^{(2,1)}(X^4)+h^{(3,1)}(X^4)-2$.
In the following we will show that the F--theory spectra eqs.(\ref{spectrumv})
and (\ref{spectrumc}) will precisely match the heterotic spectra, in particular
the number of heterotic moduli in eq.(\ref{nchiral}).

\subsection{Smooth Weierstra{\ss} models}

\vskip0.2cm

In the following we will consider F--theory on a smooth elliptically fibred 
fourfold $X^4$. So $X^4$ can be represented by a smooth Weierstra{\ss} model
without singularities. 
(We are following \cite{acl,c} in our discussion.)
So there will be no generic Non-Abelian
gauge group. We assume furthermore that generically there are no $U(1)$
factors, i.e. $r(V)=0$. However for special moduli values there might be an
enhanced gauge group with massless matter representations. However we will
not discuss this gauge symmetry enhancement for this class of models.
Since $r(V)=0$, eq.(\ref{spectrumc}) simplifies, and $N_C$ is given by
\begin{equation}
N_C={\chi\over 6}-10+2h^{(2,1)}(X^4).\label{simplnc}
\end{equation}
The Euler number of $X^4$ can be computed in terms of the topological data
of $B^3$ as \cite{sethi}
\begin{equation}
{\chi\over 24}=12+15\int_{B^3}c_1^3(B^3).\label{euler1}
\end{equation}
Moreover, since $X^3$ is assumed to be a $K3$ fibration over $\tilde B^2$, i.e.
$B^3$ is a $P^1$ fibration over $\tilde B^2$, one gets \cite{fmw}
\begin{equation}
{\chi\over 24}=12+90\int_{\tilde B^2}c_1^2(\tilde B^2)+30\int_{\tilde B^2}t^2,
\label{euler3}
\end{equation}
where $t=c_{1}({\cal T})$ (${\cal T}$ being the line bundle over $\tilde B^2$)
encodes the $P^1$ fibration structure of $B^3$. 
Note that $t$ plays the role of the $k$ of ${F_{k}}$ in the 
 string dualities in six dimensions between the heterotic string 
on $K3$ with ($12+k, 12-k$) instantons embedded in each $E_{8}$.
So in total we get
\beqa
N_C&=&38+360\int_{\tilde B^2}c_1^2(\tilde B^2)\nonumber \\ 
&+&120\int_{\tilde B^2}t^2
+2h^{(2,1)}(X^4).\label{tnc}
\eeqa

Let us now compare this F--theory computation with the number of moduli
on the heterotic side. Since $Z$ is an elliptic fibration over the same
(complex) two-dimensional base, the Hodge numbers of $Z$ are computed to be
\beqa
h^{(1,1)}(Z)=11-\int_{\tilde B^2}c_1^2(\tilde B^2),\label{hodge11z}
\eeqa
\beqa
h^{(2,1)}(Z)=11+29\int_{\tilde B^2}c_1^2(\tilde B^2).\label{hodge21z}
\eeqa
The number of gauge bundle deformations is computed from the index $I$ as
\beqa
& ~&n_{\rm bundle}=n_e-n_o+2n_o=I+2n_o\nonumber\\
&=& 16+332\int_{\tilde B^2}c_1^2(\tilde B^2)+120\int_{\tilde B^2}t^2+2n_o.
\label{nbundles}
\eeqa
We see that the number of heterotic moduli, i.e. the sum of eqs.
(\ref{hodge11z}), (\ref{hodge21z}) and (\ref{nbundles}) precisely agrees
with the number of F--theory moduli in eq.(\ref{tnc}) after setting
$h^{(2,1)}(X^4)=n_o$.

Let us briefly discuss as an example a specifical class of dual 
smooth Weierstra{\ss} models.
For this class, the (complex) two-dimensional bases $\tilde B^2$ are given
by the socalled del Pezzo surfaces
$dP_k$, the $P^2$'s blown up in $k$ points.
Then, the three-dimensional bases, $B^3$, are characterized
by  $k$ and $t$, where $t$ encodes the fibration
structure $B^3_{n,k}\rightarrow dP_k$.
Moreover we restrict the discussion to the case $k=0,1,2,3$, and $t=0$
where one has
$h^{2,1}(X^4_k)=0$, ,
$h^{1,1}(X^4_k)=3+k$, $h^{3,1}(X^4_k)=\frac{\chi}{6}-8-(3+k)=
28+361(9-k)$, $\int_{\tilde B^2}c_1^2(\tilde B^2)=9-k$. This leads to
$\chi=288+2160(9-k)$ and
\begin{equation}
N_C=38+360(9-k).\label{dpnc}
\end{equation}
On the heterotic side the Hodge numbers are $h^{(1,1)}(Z)=11-(9-k)$, 
$h^{(2,1)}(Z)=11+29(9-k)$, $I=16+332(9-k)$ and $n_o=0$. Obviously, the sum
$h^{(1,1)}(Z)+h^{(2,1)}(Z)+I$ agrees
with eq.(\ref{dpnc}).

\subsection{${\bf Z}_2$ modding of $N=2$ models}

\vskip0.2cm

Now we will use the already well established duality between four-dimensional
string compactifications with $N=2$ space-time supersymmetry to construct
new F--theory fourfolds and their dual heterotic vacua \cite{cl,acl}.  
Specifically, the
F--theory fourfolds will be ${\bf Z}_2$ modded versions of the product
space $X^3_k\times T^2$, where the threefolds are  elliptic fibrations
over the Hirzebruch surfaces $F_k$, as described in section 3. 
The ${\bf Z}_2$ modding breaks the space-time supersymmetry from
$N=2$ to $N=1$. The resulting
fourfolds are  $X^4_k=(X^3_k\times T^2)/{\bf Z}_2$.
On the heterotic side we perform the same ${\bf Z}_2$ modding, i.e.
the heterotic Calabi-Yau threefold is $Z=(K3\times T^2)/{\bf Z}_2$. In addition
we have to specify how the ${\bf Z}_2$-modding acts on the heterotic gauge
bundle which specified by the instanton number $k$.
Note that in this way we will get $N=1$ string vacua with non-trivial
gauge groups. However the action of the ${\bf Z}_2$ is such that the
charged matter field representations are non-chiral.

First, consider the heterotic compactifications. Recall that the $K3$ can 
be represented by the elliptic curve eq.(\ref{K3withWL}). The ${\bf Z}_2$
acts 
 as quadratic redefinition
on the coordinate $z$, the coordinate
of the base $P^1$ of  $K3$,
i.e. the operation is $z\rightarrow -z$. This means that
the modding is induced from the quadratic base map
$z\rightarrow \tilde z:=z^2$ with the two branch points 0 and $\infty$.
So the degrees of the corresponding polynomials $f(z)$ and $g(z)$
in eq.(\ref{K3withWL}) are reduced by half. As a result
the ${\bf Z}_2$-modding reduces $K3$  to the del Pezzo surface $dP_9$. 
This corresponds to having on K3 a Nikulin involution of type
(10,8,0) with two fixed elliptic fibers 
in the K3 leading to the following representation of $dP_9$:
\beqa
dP_9:\quad y^2=x^3-f_4(z)x-g_6(z)\label{weierdpa}
\eeqa
Representing $K3$ as a complete intersection 
in the product of projective spaces
as $K3={\scriptsize \left[\begin{array}{c|c}P^2&3\\P^1&2\end{array}\right]}$,
the ${\bf Z}_2$ modding reduces the degree in the $P^1$ variable by half; 
hence the
$dP_9$ can be represented as
$dP_9={\scriptsize \left[\begin{array}{c|c}P^2_x&3\\P^1_y&1\end{array}\right]}$.
Note that whereas the intersection lattice of $K3$ is given by
$\Lambda=E_8\oplus E_8\oplus {\bf H}\oplus {\bf H}$, the intersection lattice
of $dP_9$ is reduced to $\Lambda=E_8\oplus{\bf H}$.

Next we have to define how the ${\bf Z}_2$ modding acts on the torus $T^2$
which describes the compactification from six to four dimensions.
Namely $T^2$ will appear as an elliptic fibre over the base $P^1$ with coordinate
$z$ of $dP_9$. So the resulting space, the Calabi-Yau threefold $Z$, 
has two elliptic fibres over the same $P^1$ base, namely it has the
following fibre product structure:
\beqa
Z=dP_9\times_{P^1} dP_9.\label{hetfibrep}
\eeqa
The number of K\"ahler  deformations of $Z$ is given by the sum of
the deformations of the two $dP_9$'s minus one of the common $P^1$ base,
i.e. $h^{(1,1)}(Z)=19$. Similarly we obtain
$h^{(2,1)}(Z)=8+8+3=19$.
This Calabi-Yau 3-fold is in fact well known being one of the Voisin-Borcea
Calabi-Yau spaces. It can be obtained from $K3\times T^2$ 
by the Voisin-Borcea involution, which consists in the `del Pezzo' involution
(type (10,8,0) in Nikulins classification) with two fixed elliptic fibers 
in the K3 combined with the 
usual ``-"-involution with four fixed points in the $T^2$. 
Writing $K3\times T^2$ as $K3\times T^2={\tiny 
\left[\begin{array}{c|cc}P^2&3&0\\P^1&0&2\\P^2&0&3\end{array}\right]}$
the Voisin-Borcea involution changes this to
$Z={\tiny 
\left[\begin{array}{c|cc}P^2&3&0\\P^1&1&1\\P^2&0&3\end{array}\right]}=
dP\times _{P^1}dP$.

Finally, 
after having described the ${\bf Z}_2$ modding of $K3\times T^2$ we will now
discuss how this operation acts on the heterotic gauge bundle.
We will  consider a $N=1$ situation where after the $Z_2$
modding the heterotic gauge group still lives on a four manifold, namely on
the del Pezzo surface $dP_9$, which arises from the ${\bf Z}_2$ modding of
the $K3$ surface. In effect, the instanton numbers $n_1$, $n_2$ will be
simply reduced by half,
\beqa
l_{1,2}={n_{1,2}\over 2}.\label{nkhalfe}
\eeqa
So the total number of gauge instantons in $E_8
\times E_8$ will be reduced by two, i.e. $l_1+l_2=12$ and we are
considering $(l_1,l_2)=
(6+{k\over 2},6-{k\over 2})$ instantons in $E_8\times E_8$.
It follows that the total dimension of the instanton moduli space is reduced
by half by the ${\bf Z}_2$, which means that the quaternionic $N=2$
moduli space of $N=2$ hypermultiplet moduli  is replaced by 
a complex moduli space
of the $N=1$ chiral moduli fields.
So we see that we obtain 
as gauge bundle deformation parameters of the heterotic string on 
$Z$
the same number of massless, gauge neutral $N=1$
{\it chiral} multiplets as the number of massless $N=2$ {\it hyper} multiplets
of the heterotic string on $K3$.
This means that the $Z_2$ modding  keeps 
in the massless sector just one of the two chiral fields
in each $N=2$ hyper multiplet.
These chiral multiplets describe the Higgs phase of the $N=1$ heterotic
string compactifications.

So we can now easily determine the total number of 
heterotic chiral moduli fields as
like
\begin{equation}
N_C=38+\dim{\cal M}_{\rm inst},\label{nchetz2}
\end{equation}
where ${\cal M}_{\rm inst}$, as a function of $k$, can be read off frome table
(\ref{tab5}).

The gauge fields in $N=1$ heterotic string compactifications on $Z$
are just given by those gauge fields which arise from the compactification
of the heterotic string on $K3$ to six dimensions;  they are
invariant under the ${\bf Z}_2$ modding. 
However the complex scalar fields
of the corresponding $N=2$ vector multiplets in four dimensions
do not survive the ${\bf Z}_2$ modding. The generic gauge groups for the 
considered values of $k$ can be found in table (\ref{tab5}).
Also observe that the two vector fields, commonly denoted by $T$ and $U$,
which arise from the compactification from six to four dimensions on $T^2$
disappear from the massless spectrum
after the modding. This is expected since the
Calabi-Yau space has no isometries which can lead to massless gauge
bosons. Finally, the $N=2$ dilaton vector multiplet $S$ is reduced
to a chiral multiplet in the $N=1$ context.
At special loci in the moduli space of the $N_C$ chiral fields,
enlarged gauge groups plus charged, but non-chiral matter fields will appear.
The speciofic groups and representations are listen in \cite{acl}.

Now let us switch to the F--theory side. With respect
to $X^3_k$, the ${\bf Z}_2$ will have precisely the same action as on the
heterotic $K3$, i.e.
it acts as quadratic redefinition
on the coordinate $z_1$, the coordinate
of the base $P^1_{z_1}$ of the $K3$-fibred space $X^3_k$.
So instead of the Calabi-Yau threefolds $X^3_k$ we are now dealing
with the non-Calabi-Yau threefolds ${\cal B}^3_k=X^3_k/{\bf Z}_2$ which can
be written in Weierstrass form as follows:
\beqa
{\cal B}^3_k:\quad y^2&=&x^3+\sum_{n=-4}^4f_{4-{nk\over 2}}
(z_1)z_2^{4-n}x\nonumber \\ &+&\sum_{n=-6}^6g_{6-{nk\over 6}}(z_1)
z_2^{6-n}.\label{weierb}
\eeqa
The ${\cal B}^3_k$ are now elliptic fibrations over $F_{k/2}$ 
and still $K3$ fibrations over $P^1_{z_1}$.

Second, $X^4_k$ are of course no more products ${\cal B}^3_k\times T^2$
but the torus $T^2$ now is a second elliptic fibre which varies
over $P^1_{z_1}$.
Therefore the spaces $X^4_k$ have the form
of being the following fibre products:
\beqa
X^4_k=dP_9\times_{P^1_{z_1}}{\cal B}^3_k.\label{fibrep}
\eeqa
All $X^4_k$ are $K3$ fibrations
over the mentioned $dP_9$ surface.
The Euler numbers of all $X^4_k$'s are given by the value
\beqa
\chi=12\cdot 24=288.\label{euler}
\eeqa 
The corresponding (complex) three-dimensional IIB base manifolds $B^3_k$
have the following fibre product structure 
\beqa
B^3_k=dP_9\times_{P^1_{z_1}}F_{k/2}.\label{fibrebase}
\eeqa

The fibration structure of $X^4_k$ provides all necessary
informations to compute the Hodge numbers of $X^4_k$ from the number
of complex deformations of ${\cal B}^3_k$ in eq.(\ref{weierb}).
Without going into the details of the computations, the F--theory
result is summarized in the following table \cite{acl}:
\beqa
\begin{tabular}{|c||c|c|c|c|c|} \hline
  $k$      & $h^{(1,1)}$ & 
  $h^{(2,1)}$ & $h^{(3,1)}$ &
   $r(V)$  & $N_C$ \\ \hline\hline
0   & 12 & 112  & 140 & 0 & 262   \\ \hline
2  & 12  & 112  & 140 & 0 & 262   \\ \hline
4  & 16 & 128 & 152 & 4& 290 \\ \hline
6  & 18 & 154  & 176 & 6 & 340 \\ \hline
12  & 20 &  240 &  260 & 8 & 510 \\ \hline
\end{tabular}
\label{tab4}
\eeqa
Comparing with the corresponding heterotic numbers we get perfect agreement.

\section{Comparison of F--theory/heterotic Superpotential} 

So far we have checked the $N=1$ string-string duality for a class of models
by showing that the massless spectra match up. Now we want to go one
step further and demonstrate that for a particular dual pair also the $N=1$
superpotentials  agree. The superpotential $W$ will depend on the chiral
moduli fields, and as result of the minimization of $W$ some of the
moduli fields will be frozen to constant values. However $N=1$ space-time
will be left unbroken.

\subsection{F--theory superpotential}

\vskip0.2cm

As explained by Witten \cite{wit1}, a superpotential in F--theory
is produced by wrapping 5-branes around certain divisors $D^6\subset X^4$,
or equivalently in the IIB language by wrapping 3-branes around 4-cycles
in the base $B^3$.
Then the superpotantial is a sum over all contributing divisors like
\begin{equation}
W(z)=\sum_{D^6}\exp <c(D^6),z>,\label{suppota}
\end{equation}
where $c(D^6)$ denotes the homology class of $D^6$ and $z$ are moduli fields
in the second cohomololy of $X^4$, $z\in H^2(C^4)$.
For concreteness consider the fourfold $X^4_{k=0}=(X^3_{k=0}\times T^2)/{\bf
Z}_2=dP_9\times_{P^1}F_0$ which we discussed in section 4.2.
The corresponding type IIB base $B^3$ is simply given by the direct product
$B^3_{k=0}=dP_9\times P^1$. This implies that 
the relevant divisors $D^6$ simply have the structure
$D^6=C\times P^1\times T^2_{11-12}$, where $T^2_{11-12}$ is the elliptic
fibre of $X^4$, and $C$ is a rational curve on $dP_9$.
So to obtain a non-vanishing superpotential $W(z)$, we have 
to count all rational curves
on $dP_9$ with self-intersection $C^2=-1$. This was already done in \cite{dgw},
and the F--theory superpotential for this model has the form
\beqa
W(z_9,\tau,w_i)&=&e^{2\pi i z_9} \Theta _{E_8}(\tau,w_i)\nonumber \\
 &=&
  e^{2\pi i z_9}\sum_{a=1}^4\prod_{i=1}^8\theta_a(\tau,w_i).\label{supwi}
\eeqa
Here $\Theta_{E_8}$ is the lattice partition function of $E_8$ and
$z_9$, $\tau$ and $w_i$ ($i=1,\dots ,8$) are the moduli which correspond
to the K\"ahler structure of $dP_9$. More precisely,
viewing $dP_9$ as elliptic fibration,  $z_9$ is the K\"ahler
modulus of the base $P^1$ and $\tau$ corresponds to the modulus of the
elliptic fibre. Note that the superpotential eq.(\ref{supwi}) nicely
reflects the intersection lattice of $dP_9$, $\Lambda=E_8\oplus {\bf H}$.
Hence one can regard the $w_i$ as kind of Wilson line moduli fields.
The superpotential $W(z_9,\tau, w_i)$ is of modular weight 4 with respect to 
$PSL(2,Z)_\tau$ (the $w_i$ transform as 
$w_i\rightarrow{w_i\over c\tau+d}$).

\subsection{Heterotic superpotential}

\vskip0.2cm

As it is well known, a superpotential for (0,2) heterotic Calabi-Yau
compactifications can be either generated by world-sheet instantons or
by space-time instantons. 
Now according to \cite{wit1} the superpotential generating divisors
on the F--theory side correspond in our model to world-sheet instantons
on the heterotic side.
So we want to consider the
superpotential created by  world-sheet-instantons/rational
curves which is of the form
\begin{equation}
W(z_i)=\sum n_{d_i}\exp(2\pi id_iz_i),\label{hetsuppot}
\end{equation}
where $z_i\in H^2(Z)$ and the $n_{d_i}$ are the rational instanton numbers
of degree $d_i$.
The rational instanton  numbers of  $Z=(K3\times T^2)/{\bf Z}_2
=dP_9\times_{P^1}dP_9$  are 
essentially determined
by the $dP_9$ geometry (for more details see \cite{cl}).
Since the $dP_9$ base is common to the $F$-theory fourfolds and 
to the heterotic
$CY^{19,19}$ one can more or less immediately deduce the equality of
the superpotentials.
Specifically, the rational curved in $Z$ projected on $dP_9$ have 
self-intersection $C^2=-1$, which leads to same superpotential eq.(\ref{supwi})
as on the F--theory side.\footnote{It seems \cite{cl}
that the heterotic computation leads to a second 
$E_8$ theta-function, $\Theta_{E_8}(\tau^{\prime},w_i^{\prime})$ 
which should appear in the prefactor on the $F$-theory side.}

\subsection{Modular correction factor to $W$}

\vskip0.2cm

Now we will give some arguments that
the superpotential eq.(\ref{supwi}) has 
to be corrected by a modular function. In fact, the 
authors of \cite{dgw} expect that 
this expression for the superpotential 
has to be corrected by an $\eta(\tau)^8$ denominator - leading to a completely 
modular invariant superpotential - when taking into account a correct 
counting of the sum of rational (-1)-curves including also reducible 
objects.
We would like to argue that a different correction factor is
required to get the correct modular weight for $W$,
namely a factor $\eta(\tau)^{-12}$.
In the following we will  
give two independent arguments in favor of this modular correction
factor.

\noindent (i) So far we have done a ``naive" counting over the rational
curves of $dP_9$. However one must also include degenerate curves.
For this purpose we will use the
precise rational curve counting on the 
del Pezzo $dP_9$ provided by mirror symmetry \cite{KMV}.
If we compare the partition function of \cite{KMV} with the superpotential  
having all $w_i$ locked to zero, which is given by 
$W(\tau,w_i=0)=q_9E_4(\tau)$, we find the asserted correction factor.
In summary, we conclude that the true superpotential has the form
\beqa
W(z_9,\tau,w_i)={e^{2\pi i z_9} \Theta _{E_8}(\tau,w_i)\over \eta^{12}(\tau)}.
\label{truesup}
\eeqa
This superpotential has modular weight -2 with respect to $\tau\rightarrow
{a\tau+b\over c\tau+d}$.

\noindent (ii) The second argument in favor of this correction factor is
based on analyzing the modular transformation properties of $W(\tau)$ in
the orbifold limit of the heterotic threefold $Z=(K3\times T^2)/{\bf Z}_2$.
Namely, representing $K3=T^4/{\bf Z}_2$, $Z$ can be contructed as a ${\bf Z}_2
\times {\bf Z}_2$ orbifold, i.e. the orbifold limit of $Z$ is given by
$Z=T^6/({\bf Z}_2\times {\bf Z}_2)$. The  $T^6$ torus compactification
possesses three  moduli $T_j$ ($j=1,2,3$), which are the K\"ahler moduli
of the three subtori $T^2_j$ ($T^6=\prod_{j=1}^3T^2_j$).
As explained in \cite{cl} we can relate the orbifold moduli $T_j$ in a well
defined way to the Calabi-Yau moduli with the result that 
the modulus $\tau$ of the elliptic 
fibre of $dP_9$ corresponds in the orbifold limit  to the deformation 
along the diagonal: $\tau=T_1=T_3$. 
Now let us consider the modular transformation  properties of the effective
$N=1$ supergravity action.
In $N=1$ supergravity the K\"ahler potential $K$ and the superpotential
$W$ are connected, and the matter part of the $N=1$ supergravity Langrangian
\cite{CREMMER} is described by a single  function
$G(\phi,\bar \phi)=K(\phi,\bar \phi)+\log|W(\phi )|^2$,
where the $\phi$'s are chiral superfields.
The target space duality  transformations act as discrete
reparametrization on the scalars $\phi$ and induce simultaneously a
 K\"ahler
transformation on $K$. Invariance
of the effective action constrains $W$ to transform as a 
modular form of particular weight \cite{FLST};
specifically under $PSL(2,Z)_{T_1}\times PSL(2,Z)_{T_3}$ the 
superpotential 
must have modular weights -1. Now, with the identification $\tau=T_1=T_3$, the
K\"ahler potential at lowest order in perturbation theory is given as
$K=-2\log(\tau-\bar\tau)$. 
Then,
concerning the transformation properties 
of the superpotential under the diagonal  modular
transformations $PSL(2,Z)_\tau$,
invariance of the $G$-function requires that $W$ has modular weight -2, i.e.
that under
$\tau\rightarrow{a\tau+b\over c\tau+d}$ one has 
$W\rightarrow{W\over (c\tau+d)^2}$. So we confirm in this way the superpotential
eq.(\ref{truesup}).
Let us also remark on the
factor $q_9$ in eq.(\ref{truesup}).
 In the orbifold limit the possible 
$z_9$-depedence of the superpotential is 
again restricted by $T$-duality.
However the duality group with respect to the modulus $z_9$ is
no longer the full modular group $PSL(2,Z)$ but only a subgroup of it,
since the $R\rightarrow 1/R$ duality is broken by the freely acting
${\bf Z}_2$ in this sector. 
So the superpotential is not required to transform as a modular
function, but it should be just a periodic function in $\Re z_9$.

Finally, let us determine \cite{dgw,cl} the minimum 
of the superpotential $W(z_9,\tau,w_i)$ eq.(\ref{truesup}).
In fact, minimizing this 
superpotential leads to a supersymmetry preserving locus 
($W=0$, $dW=0$)
consisting in locking pairs of the $w_i$ on the four half-periods of the 
elliptic curve $E_{\tau}$. Expanding in $\phi_i=w_i-w_i^0$ 
around the minima $w_i^0$ 
gives $W|_{SUSY}\sim e^{2\pi iz_9}\theta_1^2(\tau,\phi)\eta(\tau)^{-6}$
behaving in leading order as 
\beqa
W|_{SUSY}\sim e^{2\pi i z_9}\phi^2,\label{wprimesusy}
\eeqa
(for notational simplicity we have identified all $\phi_i$).
The independence of $W$ from $\tau$  at the minimum is consistent with
what we expect from the orbifold limit of $Z$. We see that the superpotential
$W(z_9,\tau,w_i)$ lifts the vacuum degeneracy of this model;
the vacuum expectation values of the Wilson line
fields $\phi_{1,i}$, $\phi_{3,i}$ are set to zero after 
the minimization, i.e. the vacuum expectation values are not free,
continuous parameters in the presence of this superpotential. 
Only $z_9$ and $\tau$ survive as moduli fields. However $N=1$
space-time supersymmetry remains unbroken in the presence of the superpotential.

\section{Conclusions}

We have checked the $N=1$ string-string duality for two
classes of F--theory fourfolds compared to heterotic threefolds plus
$E_8\times E_8$ gauge bundles. The check relies on matching the
massless spectra of dual $N=1$ string pairs. In addition,
for one model 
we have supported some strong evidence that the perturbative
heterotic superpotential matches with its F--theory counterpart.
It would be very interesting to analyze models with (modular) non-perturbative,
$S$-dependent 
heterotic superpotentials \cite{sdual,sothers}  and their F--theory duals.
In this context non-perturbative symmetries in underlying $N=2$ models,
like the $S$-$T$ exchange symmetry,  may play
an important role. Also the question of breaking space-time supersymmetry
in F--theory is very important.
Moreover we would like to get a better understanding of F--theory/heterotic
dual pairs with chiral matter representations with respect to the unbroken
gauge group. 
In addition, it is a very important question, how 
transitions among $N=1$ string vacua,
possibly with a different number of chiral multiplets, take place 
\cite{n1trans}.

\vskip0.5cm
\noindent {\bf Acknowledgements}
\vskip0.2cm
\noindent I like to thank the organizers for inviting me to this very
pleasant and stimulating conference and my collaborators 
for numerous discussion on the material, presented
here.

\end{document}